\documentclass[conference,10pt,twoside, a4paper]{IEEEtran}

\usepackage{booktabs}
\usepackage{tikz}
\usepackage{amssymb}
\usepackage{graphicx}
\usepackage{amsmath}
\def\BibTeX{{\rm B\kern-.05em{\sc i\kern-.025em b}\kern-.08em
    T\kern-.1667em\lower.7ex\hbox{E}\kern-.125emX}}

\begin{document}
\title{Automated Gain Control Through Deep Reinforcement Learning for Downstream Radar Object Detection}

\author{\IEEEauthorblockN{
Tristan S.W. Stevens\IEEEauthorrefmark{1}
\qquad R. Firat Tigrek\IEEEauthorrefmark{1}
\qquad  Eric S. Tammam\IEEEauthorrefmark{2}
\qquad  Ruud J.G. van Sloun\IEEEauthorrefmark{1}\\}
\IEEEauthorblockA{\IEEEauthorrefmark{1}Department of Electrical Engineering, Eindhoven University of Technology, The Netherlands \IEEEauthorrefmark{2}Anachoic Ltd, Israel}}

\maketitle

\begin{abstract}
Cognitive radars are systems that rely on learning through interactions of the radar with the surrounding environment. To realize this, radar transmit parameters can be adapted such that they facilitate some downstream task. This paper proposes the use of deep reinforcement learning (RL) to learn policies for gain control under the object detection task. The YOLOv3 single-shot object detector is used for the downstream task and will be concurrently used alongside the RL agent. Furthermore, a synthetic dataset is introduced which models the radar environment with use of the Grand Theft Auto V game engine. This approach allows for simulation of vast amounts of data with flexible assignment of the radar parameters to aid in the active learning process.
\end{abstract}

\section{Introduction}
\label{sec:intro}

Radar is one of the key sensory systems enabling autonomous driving. A radar can retrieve radial range and velocity information of objects within the field of view of the antenna. It does so by measurement of the reflected signal which differs in frequency and phase with respect to the transmitted wave \cite{richards2014fundamentals}.

A radar system that uses information from the environment to intelligently adjust the transmitted signal is known as \textit{cognitive radar} \cite{haykin2006cognitive, greco2018cognitive}. The advantage of the cognitive radar is the ability to close the loop between receiving and transmitting elements, exploiting all relevant information through active feedback. This paper focuses on continuous automated gain control (AGC) at the cognitive radar transmitter. AGC at the receiver aims to suppress the clutter response and prevent receiver saturation \cite{Skolnik90}. AGC at the transmitter aims at low probability of intercept radar operation \cite{Shi2017}. Lower output power reduces interference to the neighboring radars in addition to saving energy. However, the probability of detection decreases with decreasing output power and targets can disappear under the noise floor. There is a clear trade-off in adjusting the transmitter gain.

Traditionally, radar detection is performed through peak detection using simple local thresholding methods such as the Constant False-Alarm Rate (CFAR) algorithm \cite{richards2014fundamentals}. These methods, however, lack the sophistication of classifying objects and clustering the resulting sparse detections as they disregard any Doppler signatures.

Deep learning methods for object detection in computer vision can be used for target detection in range-Doppler images \cite{patel2019deep, perez2019deep, major2019vehicle}. Motivated by this deep learning promise, we here extend the use to not only detection but also include optimization of radar parameters themselves using deep reinforcement learning (RL) \cite{sutton2018reinforcement}.
The proposed RL scheme is incentivized by a reward signal that is derived from the radar object detection score itself. This means that instead of optimizing the gain parameter for some surrogate measure, we directly optimize for the downstream object detection task.

We train our RL algorithm using synthetic data based on a Frequency-modulated continuous-wave (FMCW) radar model and the Grand Theft Auto V (GTA-V) game engine, which is used to model the environment. There are two main advantages to this approach. First, real radar data is expensive to acquire and label. This is a valid concern, especially since deep networks require lots of labeled data for training. Secondly, it allows the RL agent to adjust the radar parameters on the fly, and observe its impact on the environment, closing the optimization loop.
RL for radar has been introduced before, for instance for adaptive spectrum allocation \cite{Kang2018, Ma2018, liu2020decentralized, Thornton2020_2}. Our work differs from current RL applications that use higher level information to perform downstream tasks. We focus on improving the unprocessed signal using RL at the lowest level, given the raw sensory data as input to the network. Real radar data is used to qualitatively assess the object detection task and validity of the approach developed in this paper. 

In Section \ref{sec:radar_model}, we provide the radar signal model. Then, Section \ref{sec:od} describes the object detection task. The RL framework is presented in \ref{sec:rl}. Finally, the results are discussed in Section \ref{sec:results} and conclusions derived in Section \ref{sec:conclusion}.
\section{Radar Model}
\label{sec:radar_model}
\subsection{Signal processing}
\label{sec:sps}
FMCW radar is used to measure the range and speed of targets while ensuring good localization and resolution. This method uses a burst of very short chirps that ramp up in frequency. The transmitted signal is given by
\begin{equation}
    x_\text{Tx}(t) = \sum_{p=0}^{P-1}g_p(t)\exp
    \left\{
        j 2 \pi (f_c t + \frac{\alpha}{2} (t - pT)^2
    \right\},
\end{equation}
where $P$ is the number of pulses, $f_c$ is the carrier frequency, sweep rate $\alpha = B / T_s$ and
\begin{equation}
    g_p(t)=
    \begin{cases}
        1,&\text{if}~pT<t<(p+1)T\\
        0,&\text{otherwise}
    \end{cases}
\end{equation}
is the pulse function that determines the beginning and end of each sweep. The pulse duration $T_s < T$ is followed by the idle time which results in pulse repetition interval $T$. The received baseband signal with bandwidth $B$ is obtained by stretch processing $s_\text{Rx}(t) = r_\text{Rx}(t) x^*_\text{Tx}(t)$, and can be directly modeled by
\begin{equation}
    s_\text{Rx}(t) = \sum_{p=0}^{P-1}g_p(t)\exp
    \left\{
        j (\phi_1(R_0) + \phi_2(v_R) + \phi_3(R_0, p, t))
    \right\},
\end{equation}
where the dislocation of the pulse function is assumed negligible, and the phase terms are
\begin{align}
    \phi_1(R_0) &= - 2 \pi f_c \frac{2R_0}{c},\\
    \phi_2(v_R) &= -2 \pi f_c \frac{2v_R}{c} t,\\
    \phi_3(R_0, p, t) &= -2 \pi \alpha (t - pT) \left( \frac{2R_0}{c} \right),
\end{align}
which model the initial phase, Doppler shift, and delay respectively. Terms that have negligible effect on the signal phase as well as migration effects are omitted in this derivation.
The power of the reflected wave can be investigated using the radar range equation, which can be written as
\begin{equation}
    \frac{P_R R^4}{\lambda^2 \sigma} = \frac{P_T G_T G_R}{(4\pi)^3}.
\end{equation}
We refrain from making any assumptions on the antenna gains by grouping the terms on the right-hand side under a common power term $P_c$. This is the quantity used for AGC. Usually, the transmitted power is designed to ensure a certain false alarm rate and probability of detection. However, this trade-off can limit the capabilities of the radar in certain scenarios.  

The radar images in this work are constructed using a two-dimensional Fast Fourier Transform (FFT) across both fast-time and slow-time dimensions \cite{RichardsCh5}. The resulting range-Doppler image gives a representation of the raw radar signals that preserve the embedded information while allowing the use of computer vision deep learning tools.

\subsection{Data acquisition}
The radar model is built on point-clouds, where each point represents a radar scatter. In order to obtain realistic point-clouds of the environment, a game engine is used, namely Grand Theft Auto V. The data is accessed through use of the Script Hook V C++ open source library \cite{scripthookV}. The use of such a virtual environment allows for realistic dynamic settings that are consistent from one radar frame to the next.

Through means of ray casting, a detailed point-cloud can be extracted for each frame. Each point is a tuple containing the radial velocity ($v_R$), position ($R_0$), orientation, ID, and type. In this work only three kinds of objects are considered: pedestrians, vehicles and clutter. All clutter points are considered as undesired radar returns originating from stationary objects.

\begin{figure*}
    \centering
    \includegraphics[scale=0.83, trim={0cm 0.2cm 0 0},clip]{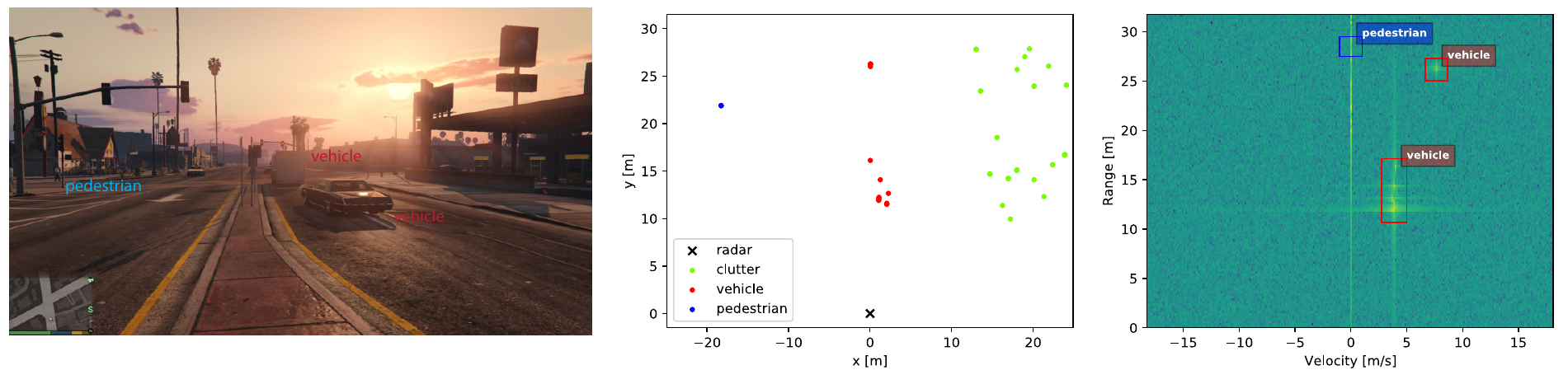}
    \caption{From left to right, the screenshot from GTA-V at the moment the data is recorded, the point cloud in Cartesian coordinates w.r.t. the radar element, and the resulting radar image produced by the FMCW model. The labeled bounding boxes for the object detection are shown as well.}
    \label{fig:pointcloud}
\end{figure*}

\subsection{Object modeling}
The behavior of reflected signals from objects can be modeled through the radar cross section (RCS) $\sigma$, which describes how much of the energy is radiated back to the source. The RCS of each object class is modeled in a different way. While there are numerous studies on the land and urban clutter statistics, e.g. \cite{Billingsley1999,Tan2008}, the radar parameters and scenarios used in this work are not covered in past research. Therefore, clutter amplitudes are drawn from a Rayleigh distribution, which gave the best fit with the clutter amplitudes in the available real data that is used for verification of the deep learning methods in this work. 

Radar returns originating from pedestrians follow a Nakagami distribution, which is obtained by fitting a distribution over the pedestrian RCS data in \cite{kamel2017rcs} along the angle of observation. Due to the limited angular resolution of the ray casting, often only a single point is assigned to a pedestrian. 

Radar returns from vehicles are modeled to depend on the vehicle orientation. The variation of RCS with respect to angle is modeled after \cite{kamel2017rcs}, using a superposition of four Sinc functions is used: wider main lobes correspond to the front and the back of the vehicle and narrower main lobes correspond to the sides. The RCS for each point scatterer on a vehicle is sampled based on the orientation and further scaled by the amount of scatters dedicated to a specific vehicle and randomized with Rayleigh distributed noise to account for the effect of multiple scattering points in a single resolution cell.


\subsection{Dataset}
The dataset consists out of 220 GTA-V scenes with 100 frames each recorded at 20 fps. 
The dataset is split into independent train and test sets. The training and test set hold 200 and 20 scenes respectively. Additionally, the training set is equally split into a part for training the RL algorithm and the object detector. 

\section{Object Detection}
\label{sec:od}
Object detection is a deep learning method where multiple objects in an image are simultaneously localized and classified. In radar target detection, this method is favored over segmentation tasks, as it can group pixels belonging to their respective targets and provide semantic information at the same time.

You Only Look Once (YOLOv3) network is a state of the art one-shot object detector \cite{redmon2018yolov3}. It is suitable for radar target detection as it has fast inference time combined with good accuracy, potentially allowing for real-time implementation of the detector. We here use Tiny-YOLO, a smaller version of the YOLOv3 network that has shown similar results to the original architecture in the context of radar detection \cite{perez2019deep}. 
For training we adopt a weighted cross-entropy (CE), which helps overcome the class imbalance that is present in the dataset. There are about twice as many vehicles compared to pedestrians, and the inverse of this ratio is used to weigh the CE loss.
Lastly, the bounding boxes in the training dataset are subjected to K-means clustering to find the proper anchor priors for the YOLOv3 architecture.
\vspace*{-5mm}
\section{Reinforcement Learning}
\label{sec:rl}
\begin{figure}
    \centering
    \includegraphics[scale=0.77]{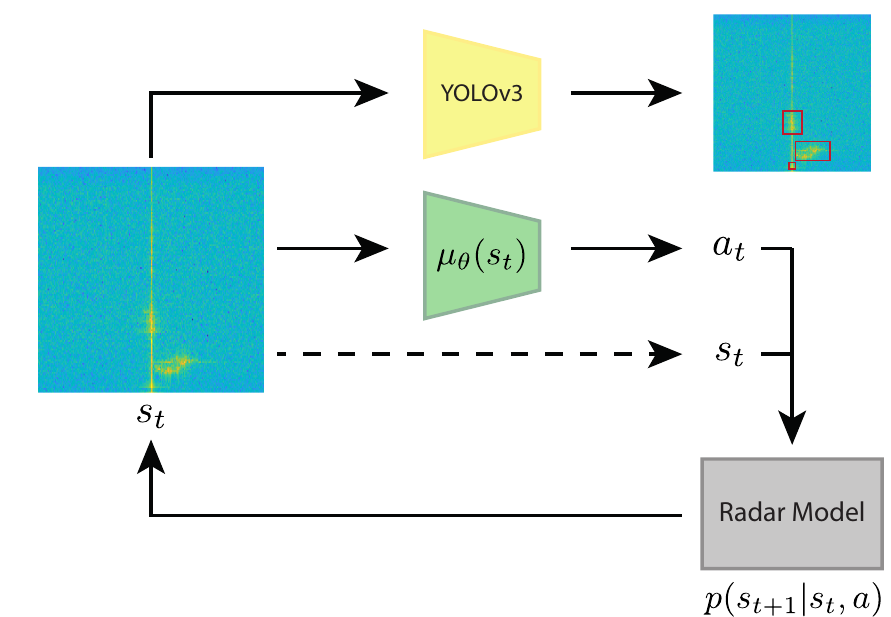}
    \caption{Inference pipeline with the RL agent $\mu_\theta$ and object detector (YOLOv3) running in parallel. The dashed line suggests implicit dependence of the next state $s_{t+1}$ with the previous state $s_t$ through the GTA-V environment.}
    \label{fig:inference_pipeline}
\end{figure}

The reinforcement learning framework is governed by an agent that takes actions according to some policy $\pi: \mathcal{S}\rightarrow \mathcal{P(A)} $ by observing a state $s\in \mathcal{S}$, which is a description of the environment. The state $s$ represents a range-Doppler image and the action $a$ corresponds to the radar power setting $P_c$. The state transitions are modeled as a Markov Decision Process (MDP). The objective of the agent is to maximize the expected cumulative reward. In this work, we use the unprocessed range-Doppler radar image directly as the input state and control the output power of the radar according to the agent's policy. The current frame will thus dictate the newly proposed radar parameter for the next frame, as is shown in Fig.~\ref{fig:inference_pipeline}. 
\vspace*{-1mm}
\subsection{DDPG}
Deep Deterministic Policy Gradient (DDPG) \cite{lillicrap2015continuous} is a model-free off-policy actor-critic network with a continuous action space that combines techniques from both "Deep Q Network" (DQN) \cite{mnih2013playing} and "Deterministic Policy Gradient" (DPG) \cite{silver2014deterministic}.

The algorithm concurrently learns two deep function approximators. An action-value function $Q_\phi^{\mu}(s, a)$, also know as Q-function, is used as the critic network parameterized by $\phi$ and describes the expected return given a state-action pair $(s, a)$, after which acting to the policy $\mu$. On the other hand, the actor network, parameterized by $\theta$ functions as the deterministic policy $\mu_\theta: \mathcal{S}\rightarrow \mathcal{A}$. 

The DDPG algorithm utilizes the Bellman equation, as many RL algorithms do, which relates the immediate reward with the expected return that can be computed by the Q-function. Learning of the critic network is performed by minimizing the mean-squared Bellman error loss through gradient descent:
\begin{equation}
    L(\phi) = \mathbb{E}_{(s_t, a_t, r, s_{t+1}) \sim \mathcal{D}} 
    \left[ 
        \big( Q_\phi(s_t,a_t) - y_t \big)^2
    \right],
\end{equation}
where $r$ is the reward function and $\mathcal{D}$ is the Replay Memory, which stores state transitions sampled from the environment. The target value $y_t$ is equal to
\begin{equation}
    y_t = r(s_t, a_t) + \gamma Q_\phi (s_{t+1}, \mu_\theta(s_{t+1})),
\end{equation}
where $\gamma=0.99$ is the discount factor. The goal is to maximize the objective function $J$, which is equal to the expected cumulative reward. This can be done by choosing actions according to the deterministic policy that maximizes the critic value. The actor network is updated through gradient ascent using the policy gradient, given by
\begin{equation}
    \nabla_\theta J \simeq \mathbb{E}_{s_t \sim \mathcal{D}} 
    \left[ 
        \nabla_\theta Q_\phi (s_t, \mu_\theta(s_t)) 
    \right].
\end{equation}
The exploration-exploitation paradigm is treated by adding noise on the action space. 
The noise is sampled from an Ornstein-Uhlenbeck stochastic process, which introduces temporally correlated exploration \cite{lillicrap2015continuous}.

\subsection{Actions and Rewards}
The output of the actor network $a$ is directly mapped to the appropriate radar power output range $P_c$ as an argument to the radar model described in section~\ref{sec:radar_model}. Unlike loss functions in supervised learning, reward functions $r(s_t, a_t)$ do not need to be differentiable. This allows use of non-smooth functions such as the object detection score in the reward formulation. During training of the RL agent, at every time-step, the F1 score of the object detector on the radar image is computed. We seek to minimize the total power output of the radar system, but not at the cost of degrading the object detection performance. To balance this trade-off, the normalized action $a_t'$ is subtracted from the F1 score as follows
\begin{equation}
    r(s_t, a_t) = \text{F1}_{\text{yolo}(s_t)} - a_t'.
    \label{eq:rewardsignal}
\end{equation}
An episode is defined as 100 consecutive frames, after which the environment is reset and a new scenery is presented. Each time-step produces a new state $s_t$ which is updated with the newly proposed power parameter but also is generated with the next point-cloud in time. This means that already during training the RL is subjected to a dynamic environment. 

\subsection{Network Architectures}
Both actor and critic networks are convolutional neural networks. The actor network has five convolutional layers with a kernel size of $3\times3$, yielding feature maps of size 32, 64, 64, 128 and 256. Each convolutional layer is followed by a batch normalization layer, ReLU activation and $2\times2$ maxpooling. The final feature map is flattened and processed by a fully-connected layer with 256 neurons, batch normalization and ReLU activation. The final layer produces the action value and has a tanh activation. 

The critic network takes both state and action as input and produces a single output value. The two-dimensional state is fed through two similar convolutional blocks, using 16 and 32 filters. Both the output feature maps of the state and the action are processed by two separate fully-connected layers with 32 neurons each. After batch normalization and ReLU activation, these two layers are concatenated and subjected to two final and similar dense layers both of size 256. The critic value is obtained after a final layer free of non-linear activation. 


\vspace*{-2mm}
\section{Results}
\label{sec:results}





\subsection{Object Detection}
\label{sec:object_detection_results}
In this section, the trained object detector is evaluated by inference on both synthetic and real radar datasets. Visually, the synthetic radar model shows similar dynamic behavior and  micro-Doppler features compared to the real data. We further validate our radar model by qualitatively assessing the object detection performance of models trained purely on simulations, when tested on real datasets.

\subsubsection{Synthetic data}
The object detection scores are generated with an intersection over union (IOU) and non-maximum suppression (NMS) threshold of both 0.5. The resulting mean average precision score is $\text{mAP}=54.8\%$.

The main difficulty for the object detection algorithm is unsurprisingly the presence of clutter along the zero Doppler axis. As a result, stationary targets, which reside on this axis, are notoriously difficult to detect. For comparison, a $\text{mAP}=74.5\%$ is obtained in the case all clutter points are removed from the dataset. These results are comparable to those found in \cite{perez2019deep}. 
\subsubsection{Real data}
\begin{figure}
    \centering
    \includegraphics[scale=0.55, trim={0cm 0cm 0 0}, clip]{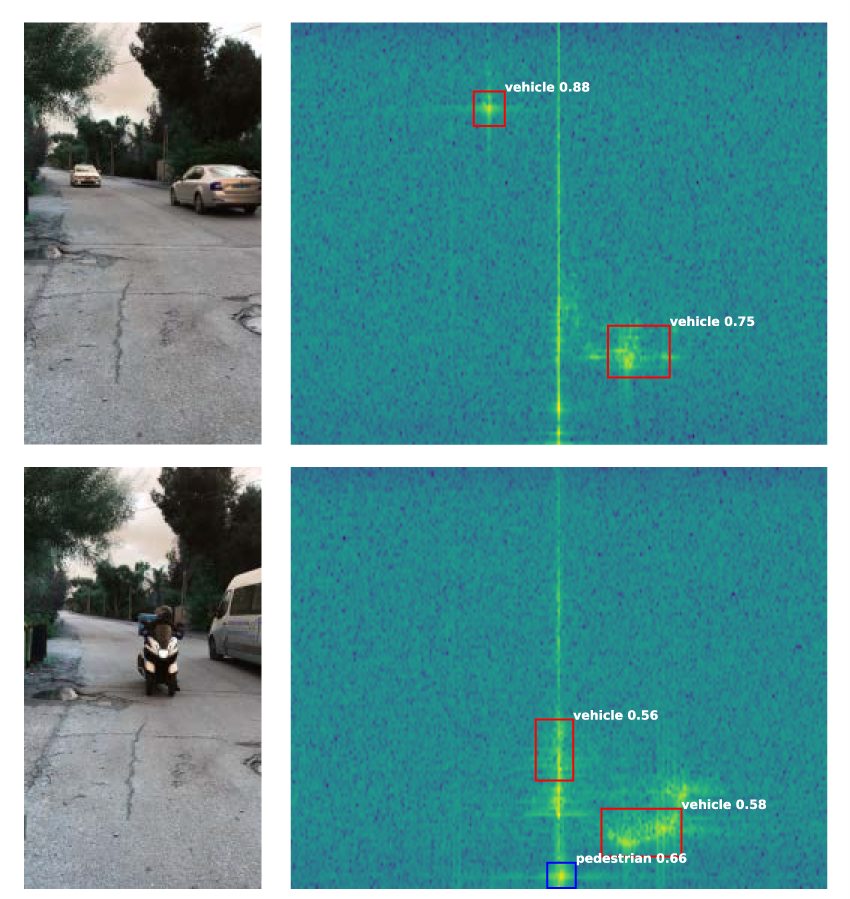}
    \vspace*{-2mm}
    \caption{Object detection inference on real radar data. The Tiny-YOLO network producing these results is trained on the synthetic dataset. The video footage is shown on the left for verification of the detection. Alongside the bounding boxes, the detection scores are given by $\Pr\{\text{Class}_i | \text{Object}\} * \Pr\{\text{Object}\}$.} 
    \label{fig:realdata}
\end{figure}

We proceed by performing inference on real radar data. The dataset consists out of 16 different videos, with a total of 13538 radar frames. 
Since our dataset is unlabeled, we qualitatively assess performance as a surrogate. Fig.~\ref{fig:realdata} shows two illustrative and typical examples. Carefull visual assessment across the entire dataset showed that all objects are tracked by the detector, meaning the objects are consistently found by the network from frame to frame. In general, visual inspection leads us to conclude that the object detection network translates well to the real radar data. 
\begin{figure*}
    \centering
    \includegraphics[scale=0.53, trim={0cm 0.4cm 0 0.4cm}, clip]{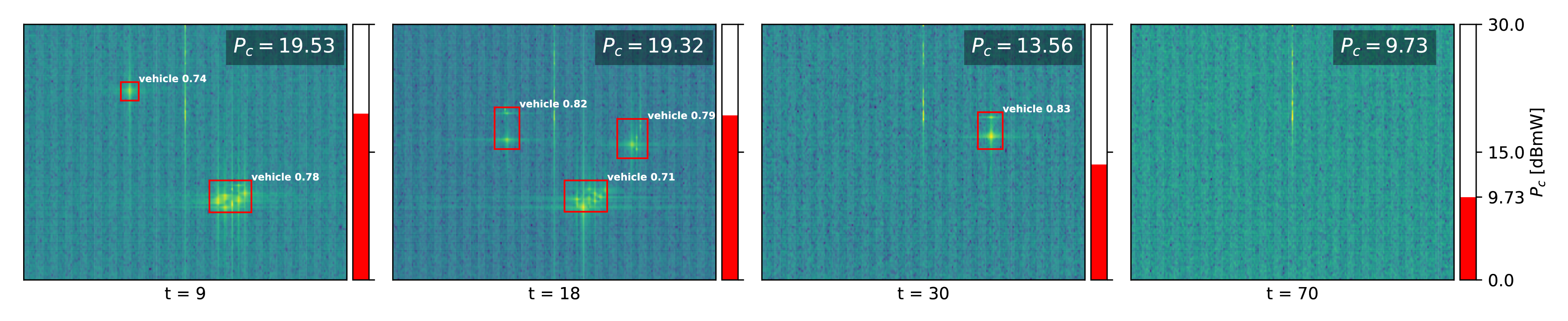}
    \caption{Four consecutive range-Doppler frames shown with detected targets by the YOLO network accompanied with the power output $P_c$ (values displayed by the red color-bar in dBmW) chosen by the RL agent. Power output is increased when more targets are present, which improves probability of detection. Power is decreased when less targets are present to reduce interference and power consumption.}
    \label{fig:results}
\end{figure*}
\begin{figure}
    \centering
    \includegraphics[scale=0.55, trim={0cm 0cm 0cm 1cm}, clip]{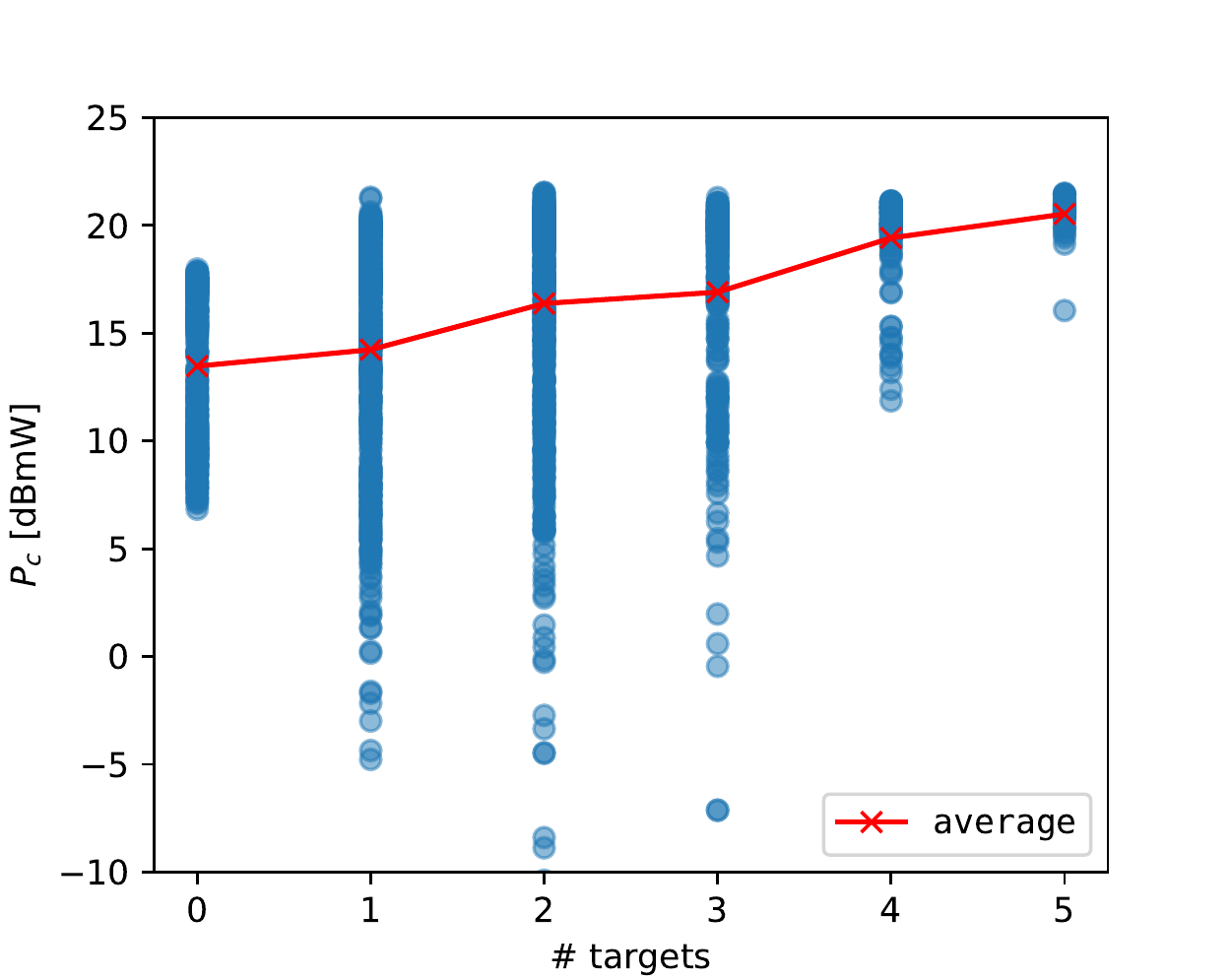}
    \caption{Average actions taken by the RL agent $P_c$ as a function of the number of targets present in each image of the test dataset. On average, the agent tends to increase the output power of the radar when more targets are present.}
    \label{fig:power_vs_ntargets}
\end{figure}
\subsection{Reinforcement Learning}
The performance of the RL agent is obtained through evaluation of the inference scheme in Fig.~\ref{fig:inference_pipeline}. The mAP score is computed over all detections made during this process. The adaptive case is compared to fixed power settings that correspond to the average actions of the agent in each scene. A mAP score increase of 1.6\% is found when RL is used compared to the non-adaptive settings.
We then investigate the behavior of the agent by analyzing action selection as a function of input states. Fig.~\ref{fig:power_vs_ntargets} shows the relationship between the number of targets present in an image and the resulting action by the agent. In general, when more targets are present the agent favors to take actions that correspond to higher output power of the radar. Additionally, the agent has recognized the cases with zero targets, and significantly reduces the power in these events. The power is never reduced to the minimum as this risks losing targets in future frames. Lastly, even though the allowed power range extends the maximum action taken by the agent by far, the agent has learned it is not necessary to increase the power further as there is no benefit of doing this with reference to the object detection task. An example of the agent in a dynamic scene is shown in Fig.~\ref{fig:results}. Visually we can observe that the agent adaptively decreases power when possible, and increases power to improve probability of detection. It is important to point out that the agent is never specifically told to adjust the power output in relation to the number of targets (number of targets is a hidden variable of the observed state), but rather has learned this through the reward signal \eqref{eq:rewardsignal} which motivates the agent to find a balance between performance and power consumption.


\section{Conclusion}
\label{sec:conclusion}
In this paper, we propose a reinforcement learning method for AGC in radar transmitter. Our method optimizes its adaptive gain policies for a given downstream task, such as radar target detection, and is trained on a dataset that combines point-cloud extraction from a game engine with a detailed FMCW radar model.

The downstream object detection task is performed by the YOLOv3 network that demonstrates good generalization towards real data, which shows the radar model and GTA environment are a viable solution for learning deep networks.

Results suggest that an RL agent is able to adaptively reduce power consumption without loss in object detection performance. In future work we will extend our framework to RL-based control of other radar parameters.


\vspace*{-1mm}
\bibliographystyle{IEEEtran}
\bibliography{refs}

\end{document}